\documentclass[
reprint,amsmath,amssymb,aps,pra,floatfix,]{revtex4-1}
\usepackage{graphicx}	% Include  files
\usepackage{dcolumn}	% Align table columns on decimal point
\usepackage{bm}		% bold math
\usepackage{color}

\begin{document}
%\preprint{APS/123-QED}

\title{Systematic effects important to 
separated-oscillatory-field measurements of
the 
$n$$=$$2$
Lamb
shift in atomic hydrogen.
}

\author{A. Marsman, 
M. Horbatsch,
Z.A. Corriveau,
and 
E.A. Hessels}
\email{hessels@yorku.ca}

\affiliation{%
Department of Physics and Astronomy, York University, Toronto, Ontario M3J 1P3, Canada
}%

\date{\today}% It is always \today, today,

\begin{abstract}
We evaluate a number of systematic effects
that are important for an experimental microwave
measurement of the 
$n$$=$$2$ 
S-to-P
intervals in atomic hydrogen.
The analysis is important for both 
re-evaluating 
the best existing measurement 
[Lundeen and Pipkin, PRL 46, 232 (1981)]
of the 
2S$_{1/2}$\nobreakdash-to\nobreakdash-2P$_{1/2}$ 
Lamb
shift,
and for a new measurement that is 
ongoing in our laboratory.
This work is part of a larger program to 
understand the 
several-standard-deviation
discrepancies between various methods for
determining the proton charge radius.

\begin{description}
\item[PACS numbers]
\verb+\pacs{32.70.Jz,06.20.Jr}+
\end{description}
\end{abstract}

\maketitle

\section{introduction}

The hydrogen 
Lamb-shift 
measurement has become 
important since it can,
when compared to very precise theory 
\footnote{for an overview of measurements of the 
hydrogen Lamb shifts and 
theoretical predictions, 
see
Refs. 
\cite{horbatsch2016tabulation}
and
\cite{mohr2016codata}.
See also 
Refs.~\cite{beyer2017rydberg}
and
\cite{fleurbaey2018new}
for two more recent measurements
in atomic hydrogen.},
determine the charge radius of the proton. 
More precise determinations of this radius have now been
performed using muonic hydrogen 
\cite{
pohl2010size,
antognini2013proton},
but there is a large discrepancy between measurements
made using ordinary hydrogen and muonic hydrogen.
This discrepancy has become known as the proton size 
puzzle 
\cite{
bernauer2014proton,
pohl2013muonic,
carlson2015proton}.

In this work, 
we evaluate possible systematic effects for a microwave 
separated-oscillatory-fields (SOF)  
precision measurement of the atomic hydrogen 
2S$_{1/2}$-to-2P$_{1/2}$ 
Lamb shift.
This interval was
measured 
by 
Lundeen 
and 
Pipkin 
\cite{PRL46.232}
in 1981,
and is currently being
remeasured by our group,
with an aim to help resolve the 
proton size puzzle.
We take advantage of the advances
in computational power that have become
available over the past decades to 
re-evaluate the 
measurement of 
Lundeen 
and 
Pipkin,
and explore the implications for 
the proton size puzzle.

In particular, 
modern computers allow for 
a full modeling of the 
microwave fields based on the 
field-plate 
geometry.
Additionally, 
modeling the time development of the 
density matrix from the time at which
the atom is created to the time of detection
is now possible.
The density-matrix calculations still require 
very intensive computations, 
and we employ the 
SHARCNET 
computer cluster.

Of particular interest for this 
re-evaluation is the possible effect of 
quantum-mechanical 
interference.
This effect requires a full 
density-matrix
modeling, 
and would not have been included
in the 
Schr$\ddot{\rm o}$dinger
equation modeling of 
Refs.~\cite{PRL46.232}
and
\cite{LundeenMetrologia}.
Such interference effects 
have been investigated by the 
present authors 
\cite{
PRA.82.052519,
PRA.84.032508,
PRA.86.012510,
PRA.86.040501,
PRA.89.043403,
PRA.91.062506,
marsman2017interference}
and by others 
\cite{
PRL.107.023001,
PRA.87.032504,
PRL.109.259901,
PhysRevA.92.022514,
PhysRevA.90.012512,
beyer2017rydberg,
PhysRevA.92.062506,
PhysRevA.94.042111,
PhysRevA.95.052503,
PhysRevA.91.030502,
AccurateLineshape}.
These investigations indicate that interference 
with a neighboring resonance,
even if it is very distant, 
can lead to significant shifts for 
precision measurements.

Also, 
we 
re-evaluate the 
AC
Stark
shifts 
and other possible systematic effects
using our modeling of the microwave
fields. 
Only small corrections are found 
relative to the 
analysis of 
Refs.~\cite{PRL46.232}
and
\cite{LundeenMetrologia}.

Finally, 
we investigate the effect 
of a possible variation of microwave
field strength as the microwave frequency
is tuned across the resonance.
We find that a correct and complete analysis
gives a smaller shift than predicted by 
Refs.~\cite{PRL46.232}
and
\cite{LundeenMetrologia}, 
and that this has a direct impact on 
the determination of the 
Lamb-shift
interval.

\section{Microwave field profile}

The microwave fields used for the 
separated-oscillatory-field measurement
of 
Ref.~\cite{PRL46.232},
employed balanced transmission lines,
as described further in 
Ref.~\cite{LundeenMetrologia}.
In 1981, 
it was not computationally
possible to simulate the full
three-dimensional,
time-dependent
fields,
and therefore a simple
two-dimensional,
DC 
model was used to 
estimate the fields.
We have simulated the fields
using the 
EMPIRE
\footnote{Empire, 
Version 4.20, 
IMST GmbH, http://www.empire.de/main/Empire/en/home.php}
software package.
This field simulation gives the 
full 
three-dimensional 
field profile, 
as well as its frequency
dependence.
Fortunately, 
the actual apparatus used to 
measure the 
Lamb 
shift has survived, 
as it was rescued from 
Harvard University
by 
Stephen Lundeen
in the early 
1990's,
and was passed on to our 
group several years ago when we
began our 
Lamb-shift
measurement.
Thus, 
our field simulations are based on 
measurements of the actual plates
that were used in the 
1981 
measurement.

\begin{figure*}
\includegraphics[width=5.5in] 
{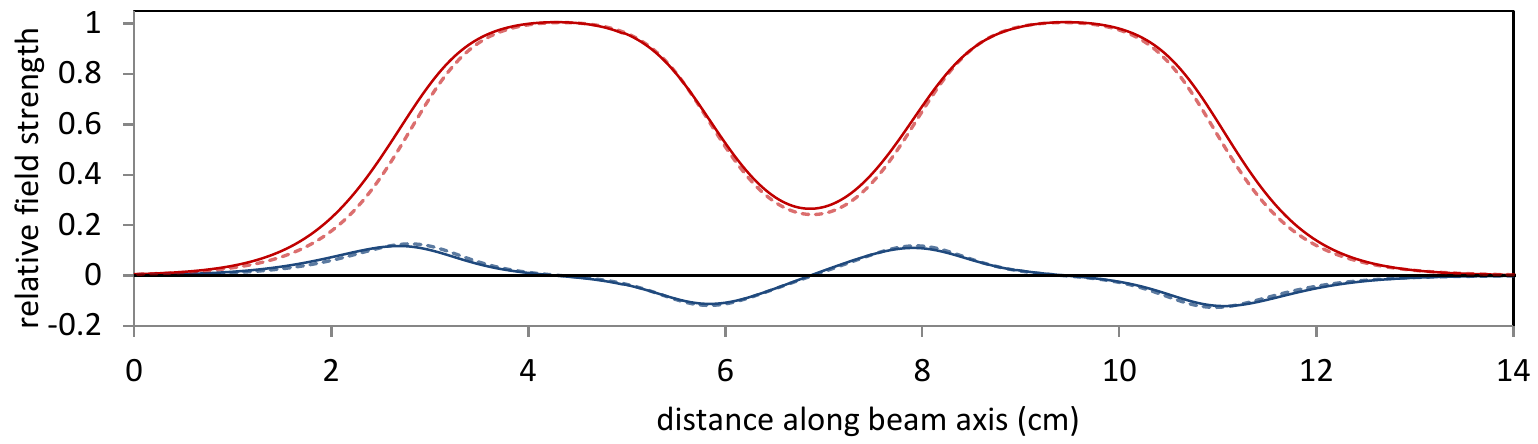}
\caption{
\label{fig:Config1Profile} 
(Color online) 
The
microwave field profile
for 
configuration~1 of 
Ref.~\cite{LundeenMetrologia}
(for the 
case where the two
SOF fields are 
in phase).
The dashed lines 
shows the field profile 
assumed in that work
(which we have recalculated 
according to their prescription),
and our simulated fields are shown
by the solid lines.
The field profiles shown are for 
a trajectory that is 
2~mm 
above the beam axis.
The larger quantity graphed
(in red)
is the vertical field component.
The smaller quantity
is the component of the field
parallel to the atomic beam.
The field profiles shown are for
910~MHz, 
but there is little variation in the 
profiles as a function of frequency.
}
\end{figure*}

That measurement used a total of 
8~configurations, 
for which the beam speed and separation between
the two SOF fields 
(between the two sets of field plates)
were varied.
We have simulated fields for all 
8~configurations.
A comparison of the fields 
used in the analysis by 
Refs.~\cite{PRL46.232}
and
\cite{LundeenMetrologia}
and our 
simulations 
for
configuration~1 
is shown in 
Fig.~\ref{fig:Config1Profile}.
The profile shown is for 
910~MHz,
and one of the concerns we had
was that this profile might vary
with frequency. 
The simulations, 
however, 
show that the profile 
varies at only the 
0.1\%
level over frequencies 
ranging between 
780
and 
1040~MHz.
The actual 
frequency-dependent
profiles are used for 
our 
density-matrix 
computations,
but the effect of the 
frequency dependence on the 
computed results is negligible.

\section{Density-Matrix calculations}

For this work, 
we perform 
density-matrix 
calculations including all
of the 
$n$=1 
and
$n$=2
states.
These include
the 
1S$_{1/2}$($f$$=$0~and~$f$$=$1),
2S$_{1/2}$($f$$=$0,~1),
2P$_{1/2}$($f$$=$0,~1),
and 
2P$_{3/2}$($f$$=$1,~2)
states,
and all 
$m_f$ sublevels,
for a total of 20 states.
Here, 
$f$ 
is the quantum number associated
with the total angular momentum
for the hydrogen atom, 
and 
$m_f$
is the projection of this angular momentum
along the quantization axis.
In a previous work, 
we extended the states to 
include higher
$n$
\cite{marsman2017interference}
and explored effects due to 
these
higher-$n$
states,
but here we restrict ourselves to just
the 
$n$=1
and 
$n$=2
states.
In total,
the density matrix
$\rho$  
has
400
elements, 
with the 
20 
diagonal elements
giving the populations,
and the 
380 
off-diagonal 
elements giving 
correlations.
The
correlations between
$n$=1
and 
$n$=2 
populations 
can safely be ignored
because the large
energy difference 
between 
$n$=1
and 
$n$=2
leads to fast oscillations 
of these elements, 
and the oscillations cause 
a cancellation when averaged
over physical processes that occur
over time scales which include
thousands or millions of these 
oscillations.

The 
density-matrix equations
follow the pattern given in 
Ref.~\cite{marsman2017interference},
and include terms due to energy differences
\begin{equation}
\label{eq:DensityMatrixEnergyTerms}
\dot{\rho}_{ba}
=
\frac{i(E_a-E_b)}{\hbar}\rho_{ba},
\end{equation}
terms due to the microwave electric field
\begin{subequations}
\label{eq:DensityMatrixFieldTerms}
\begin{eqnarray}
\dot\rho_{\rm aa}&=& 
i 
\frac{
\langle a|e \vec{E}(t)\! \cdot\! \vec{r}\ |b \rangle^*
}
{\hbar} 
\rho_{\rm ab} 
- i 
\frac
{\langle a|e \vec{E}(t)\! \cdot\! \vec{r}\ |b \rangle}
{\hbar}
\rho_{\rm ba},
\nonumber
\\
\\
\dot\rho_{\rm ba}&=& 
 i 
 \frac{
\langle a|e \vec{E}(t)\! \cdot\! \vec{r}\ |b \rangle^*
}
{\hbar}  
 (\rho_{\rm bb}-\rho_{\rm aa}) ,
\end{eqnarray}
\end{subequations}
and terms for radiative decay of populations
and correlations
\begin{subequations}
\label{eq:DensityMatrixGammaTerms}
\begin{eqnarray}
\dot\rho_{\rm aa}
&=& 
\gamma_{\rm da} \rho_{\rm dd},
\\
\dot\rho_{\rm dd}
&=&
- 
\gamma_{\rm da} \rho_{\rm dd},
\\
\dot\rho_{\rm cd}
&=& 
-\frac
{\gamma_{\rm da} + \gamma_{\rm cb}}
{2}
 \rho_{\rm cd}.
%\\
%\dot\rho_{\rm ba}&=& 
%\gamma_{\rm dacb} \rho_{\rm cd}.
\end{eqnarray}
\end{subequations}
For 
Eq.~(\ref{eq:DensityMatrixGammaTerms}),
it is assumed that states
a
and
b
are lower in energy than
states 
c
and 
d
(to allow for radiative decay),
and that 
$\gamma_{ij}$ 
are the decay rates 
from state 
$i$
to state
$j$.
All 
nonzero
terms of the form of 
Eqs.~(\ref{eq:DensityMatrixEnergyTerms}),
(\ref{eq:DensityMatrixFieldTerms}),
and 
(\ref{eq:DensityMatrixGammaTerms})
are included in the equations.
Additional terms
\cite{PRA.27.2456} 
are also included to 
account for 
quantum-mechanical 
interference between
radiative decays.

For the current work, 
we do not use the
rotating-frame
approximation, 
but simply integrate the 
time-dependent 
density-matrix
equations directly over 
the 
400-ns 
time period that it takes the atoms to
traverse the experimental 
apparatus
(from the point where the 
atoms are created by charge
exchange to the point of detection).
The 
400~ns
includes a 
35-ns-long 
preparation field at a frequency
of 
1110~MHz
(which depletes the 
2S$_{1/2}$($f$=1)
population),
the 
two 
10-ns-long
fields that form the 
SOF fields
(see, 
e.g.,
Fig.~\ref{fig:Config1Profile}),
and a 
10-ns-long
field at 
910~MHz
which quenches the 
2S$_{1/2}$($f$=0)
population by driving
it to the 
quickly-decaying 
2P$_{1/2}$
state.
These time intervals are 
reduced by a factor of 
$\sqrt{2}$
when the beam energy
is increased from 
50~keV 
to 
100~keV.
The amplitudes used for
the three microwave field regions are
26, 
11.4, 
and 
11~V/cm, 
respectively.
The radiative decay
(Lyman\nobreakdash-$\alpha$ fluorescence)
is monitored during the quenching 
field, 
and this decay is the signal 
calculated by the 
density-matrix 
equations.
This signal is calculated 
for the case when the two
SOF fields are in phase,
and when they are 
180 degrees out of phase.
The signals are also averaged
over all relative phases 
between each quench 
region and the SOF fields.
The difference between the 
in-phase
and 
180-degree-out-of-phase
signals produces the 
SOF
interference
signal that is used to measure the 
line center.
The average of these two signals
(which is referred to as 
$\bar{Q}$)
is also calculated.

\section{Quantum-mechanical interference}

The effect of 
quantum-mechanical
interference is included in the 
calculation by using the 
full 
density-matrix 
equations, 
including all terms
\cite{PRA.27.2456}
that account 
for interference between 
radiative-decay 
paths.
These additional terms have 
led to significant shifts
in a number of investigations
\cite{
PRA.82.052519,
PRA.84.032508,
PRA.86.012510,
PRA.86.040501,
PRA.89.043403,
PRA.91.062506,
marsman2017interference,
PRL.107.023001,
PRA.87.032504,
PRL.109.259901,
PhysRevA.92.022514,
PhysRevA.90.012512,
beyer2017rydberg,
PhysRevA.92.062506,
PhysRevA.94.042111,
PhysRevA.95.052503,
PhysRevA.91.030502,
AccurateLineshape}
of 
quantum-interference 
effects on precision measurements.
The effect of this quantum 
interference cannot be accounted for
using an integration of the 
time-dependent 
Schr$\ddot{\rm o}$dinger
equation, 
and therefore could not be
invesigated by the analysis of 
Refs.~\cite{PRL46.232}
and
\cite{LundeenMetrologia}.

We have intentionally included the 
2P$_{3/2}$ 
states
in our analysis, 
even though 
the 
2S$_{1/2}$-to-2P$_{3/2}$
transition is
9~GHz 
out of resonance with
our microwave fields.
An analysis
\cite{PRA.86.012510}
 of a similar 
SOF 
measurement in the 
2$^3$P 
states of helium
showed that 
quantum-mechanical shifts 
can be caused by even
very far
off-resonant 
transitions.
Here, 
however,
we see no 
quantum-interference
shifts. 
That is, 
we observe identical line centers 
with or without the inclusion of the
interference terms in the 
these density-matrix 
equations. 
We note that our earlier work 
\cite{marsman2017interference}
did
show interference effects 
when 
higher-$n$
states are included,
but these shifts were due
to a processes that involved 
interferences between,
for example,
2S-to-2P 
transitions 
and 
3S-to-3P
transitions,
and resulted from the fact 
that these two 
$n$$=$3
states can radiatively
decay down to the two 
$n$$=$2 states.

\section{Calculated shifts}

We have used the 
density-matrix 
equations,
along with our simulated 
microwave fields,
to calculate 
the lineshapes for 
each of the 
8~configurations
used in the 
Lundeen 
and 
Pipkin 
measurement.
We have performed the 
simulation at the same set
of frequencies as were 
used in the
measurements,
and have determined
the line center
in the same 
manner as was used
in the experiment.
The shifts that we observe
in our simulations are separated
into four categories. 
We first calculate the shifts that 
occur for atoms that are traveling 
along the central axis of the experiment,
while also making the simplifying assumption
that the 
2S$_{1/2}$($f$=1) 
states have no initial population.
The shifts that we obtain are shown in 
row~2 
of 
Table~\ref{tbl:shifts},
and are compared 
to the similar 
shifts obtained in 
Ref.~\cite{LundeenMetrologia}.
We find some small differences
(of between 
1 
and 
9~kHz)
between our
and their analysis
for the 
8~configurations. 
However,
the weighted average difference,
using the weights shown in row~1
of the table
(which are the weights used in
Ref.~\cite{LundeenMetrologia}),
differs by only 
-2.6~kHz,
only marginally larger than the 
uncertainty that they assigned to 
this systematic effect.

\begin{table*}
\centering
\caption{
\label{tbl:shifts} 
A comparison of shifts predicted in this work 
to those found in 
Table~10
of
Ref.~\cite{LundeenMetrologia}
for the 
2S$_{1/2}(f$$=$$0)$-to-2P$_{1/2}(f$$=$$1)$
interval in atomic hydrogen.
The shifts are in kHz, 
with uncertainties in the last digits
in parentheses. 
Shifts are calculated for all
8~configurations used in 
Ref.~\cite{LundeenMetrologia},
and weighted averages for the 
8~configurations
are also calculated. 
}
\begin{tabular}{lllllllllll}
\\
\hline
\hline
row&
systematic effect
&cfg.1
&cfg.2
&cfg.3
&cfg.4
&cfg.5
&cfg.6
&cfg.7
&cfg.8
&{\bf wt. avg.}
\\
\hline
\\
1&
{weight assigned by \cite{LundeenMetrologia}}&
{\it 0.069}&{\it 0.090}&{\it 0.218}&{\it 0.090}&
{\it 0.033}&{\it 0.2955}&{\it 0.182}&{\it 0.0225}
\\ 

\\
2&
f=0 only; level power; on-axis&
38.6&27.0&26.0&22.0&18.4&25.1&20.3&18.1
&{\bf 24.9}\\

3&
cf. Table~10 row 2) \cite{LundeenMetrologia}&
38(4)&36(2)&30(3)&24(3)&21(3)&27(2)&21(2)&19(2)
&{\bf 27.5(2.5)}\\
4&
difference&
0.6&-9.0&-4.0&-2.0&	-2.6&-1.9&-0.7&-0.9
&{\bf -2.6}
\\		
\\
5&
additional off-axis shift&
 6.2& 5.2& 5.2& 3.9& 3.6& 4.1 &	3.7 & 2.9
 &{\bf 4.4}\\
6&
cf. Table~10 row 6) \cite{LundeenMetrologia}&
6(4)&3(2)&3(2)&3(2)&3(2)&3(2)&2(1)&2(1)
&{\bf 3.0(1.9)}
\\
7&
difference&
0.2&2.2&2.2&0.9&0.6&1.1&1.7&0.9
&{\bf 1.4}
\\ 
\\
8&
additional $f$=1 shift&
 0.1& 0.0& 0.0& 0.0& 0.0& 0.1&-0.3& 0.1
& {\bf 0.0}\\
9&
cf. Table~10 row 4)  \cite{LundeenMetrologia}&
4(1)&2(1)&0(0)&0(0)&0(0)&-1(0)&0(0)&0(0)
&{\bf	0.2(0.2)}
\\
10&
difference&
-3.9&-2.0&0.0&0.0&0.0&1.1&-0.3&0.1
&{\bf -0.2}
\\
\\
11&
$-$0.11\%/100 MHz field slope&
-2.3&-0.8&-0.3&-0.2&-0.1&-0.2& 0.0&-0.2
&{\bf -0.4}\\
12&
cf. Table~10 row 7)  \cite{LundeenMetrologia}&
-6(6)&-4(2)&-1(1)&-1(1)&0(0)&1(1)&0(0)&0(0)
&{\bf -0.8(1.2)}
\\
13&
difference&
3.7&3.2&0.7&0.8&-0.1&-1.2&0.0&-0.2
&{\bf 0.4}
\\
\\
14&
total corrections (this work)&
45.0&	32.2&	31.2&	25.9&	22.0&	29.2&	23.8&	21.0
&{\bf 29.3}\\
15&
total (row 2), 6), 4), 7)  \cite{LundeenMetrologia})&
42(8)&37(4)&32(4)&26(4)&24(4)&30(3)&23(2)&21(2)
      &{\bf 29.9(3.4)}
\\
16&
difference&
0.7&-5.6&-1.1&-0.3&	-2.1&-0.9&0.7&-0.2
&{\bf	-0.9}
\\
\\
17&
final centers with our corrections\\ 
&(minus 909 MHz)&
926(10)&895(12)&911(8)&873(14)&	875(23)&
893(11)&887(14)&871(40)
\\
\\
18&
$b$=0.5(2)\%/100~MHz field slope&
10.4(4.2)&3.5(1.4)&1.2(0.5)&1.0(0.4)&0.6(0.2)
&0.8(0.3)&0.0(0.0)&0.9(0.4)
&{\bf 1.7(0.7)}\\
19&
cf. Table~10 row 8)  \cite{LundeenMetrologia}&
25(10)&18(7)&9(3)&6(2)&2(1)&5(2)&2(1)&2(1)
&{\bf 7.8(3.0)}\\
20&
difference&
-14.6&-14.5&-7.8&-5.0&-1.4&-4.2&-2.0&-1.1
&{\bf	-6.1}\\
\\
21&
final centers with our corrections\\
&including row 18 (minus 909 MHz)&
916(11)&891(12)&910(8)&872(14)&
874(23)&892(11)&887(14)&870(40)
\\
22&
Our suggested corrected result&
&&&&&&&&
{\!\!\!\!\!\!\!\!\!\!\!\!\!\!\bf 909.894(20)~MHz}
\\
\\
\hline
\hline
\end{tabular}
\end{table*}

The corrections for the fact that 
some atoms are displaced from the central 
axis of the beam line as they 
pass through the 
SOF
fields
has also been
calculated using the full 
density-matrix method.
The 
off-axis
trajectories could be important,
since,
as shown in 
Fig.~\ref{fig:Config1Profile},
they 
(unlike the on-axis trajectories)
have components of the microwave
electric field along the direction of the
beam axis.
As predicted in 
Ref.~\cite{LundeenMetrologia}),
the correction depends quadratically 
on how far the atoms are away from 
the central axis,
and using their estimate that the 
rms 
distance is 
1.265~mm, 
we obtain the corrections shown in 
row~5
of  
Table~\ref{tbl:shifts}.
These agree within the stated
uncertainties with the similar corrections
in Ref.~\cite{LundeenMetrologia},
and,
as shown in the table,
the weighted average of 
our corrections 
and theirs
differs by only 
-1.4~kHz.

Similarly, 
the corrections due to 
the inclusion of 
initial populations in the 
$f$=1
states
leads to shifts 
(row 8)
that are in reasonable agreement
with 
Ref.~\cite{LundeenMetrologia}.
The difference in the weighted 
average for this additional effect
is completely negligible 
at only 
-0.2~kHz.

The final correction that we calculate is
that due to a possible variation of microwave
power as the microwave frequency is tuned
across the resonance.
The measured value for this field slope
in 
Ref.~\cite{LundeenMetrologia}
is 
$-$0.11(12)\%/100~MHz.
Our calculations for this effect are shown
in 
row~11 of the table.
Although our corrections agree to within 
the stated uncertainties for this effect
calculated in
Ref.~\cite{LundeenMetrologia}
(shown in 
row~12),
our corrections are systematically smaller by
about a factor of three.
The agreement between our corrections 
and theirs is due only to the large 
(more than 
100\%)
uncertainty in their measured 
field slope.

In 
row~14, 
we show the total of all four effects discussed.
These are compared to the similar
totals from 
Ref.~\cite{LundeenMetrologia} 
in 
row~15.
The corrections
agree to within the 
uncertainties given in the original work,
and the weighted average of our total
correction agrees with theirs to better 
than 
1~kHz.
The final line centers for each configuration
(using our corrections) 
is shown in 
row~17
of the table. 
The results for the 
8~configurations
are not in good agreement,
with a standard deviation of 
18~kHz,
which is much larger than the 
uncertainties for most of the 
configurations.

\section{Additional field slope}

Lundeen and Pipkin resolved the
discrepancies between the centers 
for different configurations
by noting that there were three lines
of evidence that pointed to a larger
than expected 
field slope,
with an additional slope of
$b$=+0.5(2)\%/100~MHz
(over and above the field slope
directly measured).
Based on these three lines of 
evidence, 
they add an 
{\it ad hoc}
correction to their line centers
based on 
$b$.

The first, 
and most statistically significant,
evidence for the existence of 
an additional field slope 
$b$
came from an analysis of the 
asymmetry 
(about the 
SOF 
line center)
of the 
$\bar{Q}$
signal.
Our analysis of their
observed asymmetry 
agrees with their 
assessment of the implied 
$b$
(as listed in 
column~2 in 
Table~14 
of
Ref.~\cite{LundeenMetrologia}).
In particular, 
we verify 
their 
un-numbered
equation near the top
of 
column~2 
of 
page~47,
which shows the scaling 
of 
$\bar{Q}$
with field strength,
and which they used to 
determine 
$b$.

The second line of evidence
that they used to show that 
an additional field slope 
$b$ 
was present
is based on the accompanying 
equation on 
page~47
for the scaling with 
field strength of the
SOF
interference signal
$I$. 
In this case, 
our analysis shows that this
scaling does not correctly
predict the effect on the 
SOF
signal
due to an increase or decrease
in field strength,
and,
therefore,
does not properly calculate
the asymmetry
caused by a field slope.
Thus, 
the results in the third
column of 
Table~14 
of
Ref.~\cite{LundeenMetrologia}
give incorrect estimates of 
$b$.

The third line of evidence that 
Lundeen
and 
Pipkin 
used 
involves the 
SOF 
line centers themselves.
They make the case that the inclusion
of the larger slope
makes these line centers more consistent.
We show the corrections that they applied
in row~19
of Table~\ref{tbl:shifts}. 
However, 
because the scaling
formula that they used does not 
agree with our analysis, 
we get a different
(and about a factor of three smaller) 
effect due to a 
$b$=0.5\%/100~MHz
field slope, 
as shown in 
row~18 
of the table.
The weighted 
average of the shifts 
differ by  
$6.1$~kHz,
as shown in 
row~20.
More importantly,
our shifts indicated in 
row~18
of 
the table
do not serve to resolve the 
discrepancies between the 
line centers obtained for the 
8~configurations,
as shown in 
row~21.
The 
8~centers 
corrected for 
$b=$0.5\%/100~MHz
do not show acceptable agreement 
with each other, 
and
still have a standard deviation
of 
16~kHz.

\section{Corrected Center}

Given our 
re-analysis 
of the systematic effects
we take the weighted
average of the values in 
row~21 
of 
Table~\ref{tbl:shifts}
to be the best estimate of the 
center measured in that work.
The only significant disagreement
that we find with the original analysis
is the treatment of the field slope,
and,
in particular,
the 
{\it ad hoc}
correction due to an additional
field slope.
Our 
re-analysis 
of the effect of 
$b$=0.5\%/100~MHz 
leads to a 
shift of 
$-$6~kHz in the resonance 
center 
(row 20),
which
(along with another 
$-$1~kHz shift from all other effects,
as shown in row 16)
changes the measured value for the
2S$_{1/2}$($f$$=$0)-to-2P$_{1/2}$($f$$=$1)
interval in atomic hydrogen
from 
909.887~MHz
to
909.894~MHz.

Four considerations need to be included in the 
assessment of 
the uncertainty in this measurement.
The first consideration is the 
9~kHz
uncertainty from the original analysis
of this measurement.
The second consideration is the uncertainty 
as to whether any level
of correction should have been made 
for the 
{ad hoc} additional slope $b$.
Given two of the lines of evidence 
indicated by 
Lundeen 
and 
Pipkin 
do
not hold up in our 
re-analysis, 
it may not make sense to still make a
correction.
Since our analysis gives a correction of only
1.7~kHz for 
the additional field slope 
$b$,
we assign an additional uncertainty of 
1.7~kHz.
The third consideration is the assignment 
of an uncertainty is the fact that with our 
re-analysis 
the centers in the 
8~configurations no longer agree with each 
other.
The standard deviation for the 
8~configurations 
is 
16~kHz
(or 
18~kHz
if we do not make an 
{\it ad hoc}
correction for 
$b$).
We assign an additional uncertainty
equal to this standard deviation,
since it is not clear,
without
identifying a cause for the discrepancy,
which of the configurations gives the 
correct center for the interval.

When added in quadrature, 
the uncertainty from these three sources
gives 
20~kHz,
and is dominated by the 
standard deviation of the 
8 configurations.
A final consideration is found in our 
previous 
work \cite{marsman2017interference},
in which we showed that the effect 
of 
higher-$n$
states could be of the order of 
10~kHz.
Since the 
20~kHz uncertainty assigned here
is larger than 
the 
this 
10-kHz
scale,
and since,
as discussed in 
Ref.~\cite{marsman2017interference},
we cannot calculate the 
actual shift caused by 
higher-$n$,
we do not assess any additional uncertainty
due to the possible effect from these
states.

The final estimate for our 
re-evaluation 
of the 
measured value for the
2S$_{1/2}$($f$$=$0)-to-2P$_{1/2}$($f$$=$1)
interval is
therefore
909.894(20)~MHz.
This value agrees 
(to within its uncertainty) 
with the original value of 
Refs.~\cite{PRL46.232}
and
\cite{LundeenMetrologia}.
It also agrees
(to within its uncertainty)
with 909.8742(3),
which is 
the value
\cite{horbatsch2016tabulation}
that is 
predicted from precise 
quantum-electrodynamics
theoretical predictions for the 
Lamb shift, 
along with the proton charge
radius obtained from muonic hydrogen
and contributions from the hyperfine
structure.
The larger uncertainty assigned here
reduces the strength of the measurement
for determining the proton charge radius,
and the result is consistent with both the 
larger radius suggested by 
CODATA
\cite{mohr2016codata}
and the smaller radius 
suggested by some recent
measurements
\cite{beyer2017rydberg,
pohl2010size,
antognini2013proton}.

\section{conclusions}

We have 
re-analyzed the 
$n$$=$$2$ 
Lamb-shift
measurement made in 
1981 by Lundeen and Pipkin,
taking advantage of 
computing improvements in the 
intervening decades.
Using a more sophisticated analysis
of both the fields and of the atomic
physics processes,
we find that we largely agree with
their original analysis.
We find, 
however, 
one correction to their analysis.
In all,
we suggest that their result should be
corrected by 
7~kHz,
and their uncertainty be increased to 
20~kHz.
The increased uncertainty makes the 
measurement consistent with both the 
small and large values of the proton 
charge radius that have been the subject
of the recent proton radius puzzle.
The analysis performed in this work will 
directly apply to a new measurement of the 
same interval that is
on-going in our laboratory.

\section{acknowledgements}
This work is supported by 
NSERC 
and a 
York Research Chair 
and used 
SHARCNET 
for computation.

\bibliography{LunPipInterference}

\end{document}